\documentclass[fleqn,usenatbib]{mnras}

\usepackage{newtxtext,newtxmath}
\usepackage[T1]{fontenc}
\usepackage{ae,aecompl}
\usepackage{graphicx}
\usepackage{amsmath}
\usepackage[mathscr]{euscript}

\let\mathscr\relax % so we can load this and rsfs
\usepackage[scr]{rsfso}
\usepackage{enumitem}
\usepackage{tabularx, booktabs, makecell, caption}
\usepackage{siunitx}
\newcommand{\daa}{\Delta\alpha/\alpha}

\DeclareMathAlphabet{\mathpzc}{OT1}{pzc}{m}{it}

\title[AI-VPFIT]{Artificial intelligence and quasar absorption system modelling; application to fundamental constants at high redshift}

\author[C. C. Lee et al]{
Chung-Chi Lee$^{1}$\thanks{E-mail: lee.chungchi16@gmail.com},
John K. Webb$^2$\thanks{E-mail: jkw.phys@gmail.com},
R. F. Carswell$^{3}$,
and
Dinko Milakovi{\'c}$^{4}$.\\\\
$^{1}$DAMTP, Centre for Mathematical Sciences, University of Cambridge, Cambridge CB3 0WA, UK\\
$^{2}$Clare Hall, University of Cambridge, Herschel Rd, Cambridge CB3 9AL.\\
$^{3}$Institute of Astronomy, University of Cambridge, Madingley Road, Cambridge CB3 0HA, UK\\
$^{4}$European Southern Observatory, Karl-Schwarzschild-Str. 2, 85748 Garching bei M\"unchen, Germany
}

\date{Accepted April Fools Day 2021.}

\pubyear{2021}

\begin{document}
\label{firstpage}
\pagerange{\pageref{firstpage}--\pageref{lastpage}}
\maketitle

\begin{abstract}
We have developed a new fully automated Artificial Intelligence-based method for deriving optimal models of complex absorption systems. The AI structure is built around VPFIT, a well-developed and extensively-tested non-linear least-squares code. The new method forms a sophisticated parallelised system, eliminating human decision-making and hence bias. Here we describe the workings of such a system and apply it to synthetic spectra, in doing so establishing recommended methodologies for future analyses of VLT and ELT data. One important result is that modelling line broadening for high-redshift absorption components should include both thermal and turbulent components. Failing to do so means it is easy to derive the wrong model and hence incorrect parameter estimates. One topical application of our method concerns searches for spatial or temporal variations in fundamental constants. This subject is one of the key science drivers for the European Southern Observatory's ESPRESSO spectrograph on the VLT and for the HIRES spectrograph on the ELT. The quality of new data demands completely objective and reproducible methods. The Monte-Carlo aspects of the new method described here reveal that model non-uniqueness can be significant, indicating that it is unrealistic to expect to derive an unambiguous estimate of the fine structure constant $\alpha$ from one or a very small number of measurements. No matter how optimal the modelling method, it is a fundamental requirement to use a large sample of measurements to meaningfully constrain temporal or spatial $\alpha$ variation.
\end{abstract}

\begin{keywords}
Cosmology: cosmological parameters; Methods: data analysis,  numerical, statistical; Techniques: spectroscopic; Quasars: absorption lines; Line: profiles; Abundances
\end{keywords}

\section{Introduction}\label{sec:intro}

The development of optimal, objective, and reproducible methods for the analysis of high-resolution absorption systems is essential for analysing the high quality data now being obtained on new astronomical facilities. Forthcoming facilities like the ELT clearly demand sophisticated analytic tools. Searches for new physics, revealed by temporal or spatial variations in fundamental constants, constitute one of the three main science drivers for the ESPRESSO spectrograph on the European Southern Observatory's VLT \citep{Pepe2014}, and one of the key goals for the forthcoming ELT \citep{Hook2009,ESO_ELTbook2010,ESO_ELTbook2011,ELT_Liske2014,Marconi2016}.

Searches for variations in fundamental constants in the high redshift universe can be carried out using quasar spectroscopy. Gas clouds from galaxies falling sufficiently near the Earth-quasar sightline impose absorption lines on the background quasar. Careful modelling of these absorption lines allows us to construct a detailed picture of the physics in these systems. The observational data may include a large number of absorption transitions from many elements at a range of ionisation states. Moreover, the kinematic structure of the absorbing gas can be highly complex and each system may comprise a large number of absorbing components, each at a slightly different redshift. For these reasons, the required number of model parameters can be large and the analysis tools must be sophisticated.

\cite{Bainbridge2017,gvpfit17} first developed AI methods that were capable of modelling arbitrarily complex absorption systems without any human interactions. At the heart of that process is VPFIT \citep{VPFIT,web:VPFIT}, a non-linear least squares system with tied parameter constraints\footnote{We note there are several other codes also designed to fit absorption complexes. Amongst these are FITLYMAN \citep{Fontana1995}, VIPER \citep{Gaikwad2017}, ALIS \citep{web:Cooke2015}, TRIDENT \citep{Hummels2017}, and VoigtFit \citep{Krogager2018}. We additionally note that VIPER has automated capabilities and that VPFIT is able to perform automated Lyman-$\alpha$ forest fitting.}. Optimal measurements of the fine structure constant, $\alpha \equiv e^2/(4 \pi \epsilon_0 \hbar c)$, make use of the {\it Many Multiplet Method} \citep{Dzuba1999,Webb1999} which is explicitly incorporated into VPFIT, such that $\alpha$ can be directly measured within the non-linear least squares procedure as one of many free parameters describing each absorption complex. The mechanism for doing this is described in the VPFIT documentation \citep{web:VPFIT}.

In the Bainbridge code and in the new methods introduced in this paper, the additional genetic code replaces previously required human interactions, guiding the descent direction to an optimal solution. In order to select an ``appropriate'' model, by which we mean a model having the ``right'' number of free parameters, we use an information criterion (IC). ICs have the general form
\begin{equation}
    IC = \chi^2 + \mathrm{penalty \ term} 
\end{equation}
where the ``penalty term'' is a function that depends on the number of data points and the number of free parameters in the model. Different ICs have different penalty terms (e.g. the corrected Akaike Information Criterion has a penalty term that is generally smaller than the Bayesian Information Criterion). In a separate detailed study \citep{webb2020getting}, we compare the performances of these two ICs and also introduce a new IC, ``SpIC'', designed specifically for spectroscopy, that offers some advantages over the well-known AICc and BIC. However, in order to keep the present paper as simple as possible, we opt to use only AICc in this paper. 

The system introduced in this paper makes simultaneous use of genetic algorithms and non-linear least squares and is capable of carrying out complex tasks autonomously that would otherwise be performed by intelligent beings. The method is thus appropriately called ``artificial intelligence'', which we name ``AI-VPFIT''. Hybrid algorithms that unify AI methods with non-linear least squares techniques have also been referred to as ``memetic algorithms'' \citep{Moscato1989}.

In the present paper, we develop these ideas further, the key aims being to find faster algorithms and improve the overall performance, particularly for complex absorption systems/datasets that may be challenging to model, comprising a wide range of line strengths, from heavily saturated to barely detected, allowing for any blended absorption lines from other systems.

Performance checking focuses on estimating the fine structure constant in quasar absorption systems, because this is a particularly challenging problem, more so than simple 3-parameter absorption line fitting. 
We parameterise a change in the fine structure constant as $\Delta\alpha/\alpha \equiv (\alpha_z - \alpha_0)/\alpha_0$, where $\alpha_z$ is the value measured at absorption redshift $z$ and $\alpha_0$ is the terrestrial value of $\alpha$. Whilst the newly created methods introduced in this paper have been developed in the context of quasar spectroscopy, they are likely to be useful in any absorption line spectroscopy application, including stellar spectroscopy and terrestrial experiments.

In Section \ref{sec:AI} we explain the various stages of the AI-VPFIT algorithm. We partition the process into 6 stages, describing the design of each in detail. Section \ref{sec:synthetic} evaluates the performance of AI-VPFIT using synthetic spectra. Section \ref{sec:Temp-importance} shows that it is important to properly model the absorption line width and that not doing has a significant impact on parameter estimation. In Section \ref{sec:nonuniqueness}, we discuss how an objective AI method removes biases inherent to previous interactive analyses and show that model non-uniqueness limits the accuracy achievable from a single measurement of variation in the fine structure constant $\daa$. Extensive use is made of the corrected Akaike Information Criterion for comparing the relative improvement in model development at each stage. The pros and cons of this are discussed in Section \ref{sec:Overfitting}. Finally, in Section \ref{sec:conclusions} we summarise AI-VPFIT's key characteristics and performance, the main findings from extensive testing using synthetic data, and the implications and requirements of future measurements to constrain $\daa$ using quasar spectroscopy.

\section{The AI Method}\label{sec:AI}

The new algorithms described in this paper follow the broad principles outlined in \cite{Bainbridge2017,gvpfit17} (GVPFIT) although there are notable differences and many new algorithms. These include: (i) as described in the following text, trial line positions are random; (ii) the ordering of the various tasks carried out here is different here compared to \cite{Bainbridge2017,gvpfit17}; (iii) Bayesian model averaging is not used here; we next describe each stage of the modelling process; (iv) GVPFIT was set up to use either fully turbulent or fully thermal broadening. This turns out to be a significant issue (discussed in detail later in this paper). The new method presented here does not make these assumptions; (v) two new stages are included to refine the model and address the overfitting problem.

Throughout this paper we will necessarily be using various terminologies. The meaning of each term is expanded on later in the text but for clarity we provide brief definitions here:
\begin{itemize}[leftmargin=0.5cm]
\item {\it Target line}: this is a heavy element transition forming part of the absorption complex being fitted, i.e., the redshift system(s) of interest. A target line can be primary or secondary. Throughout the paper we may also refer to this as a {\it velocity component};
\item {\it Primary transitions} and {\it primary species}: we designate one or more transition (from the same species) for the purposes of generating a preliminary velocity structure for the absorption system being modelled. The primary species plays an additional important role in that it becomes the species to which secondary species' absorption line parameters are tied\footnote{For VPFIT users: the {\it primary species} in AI-VPFIT takes lower case letters in the VPFIT first-guess input file.}.
\item {\it Secondary transitions/species}: these are the additional transitions (from a different element or ion) brought into the model after the preliminary model has been made;
\item {\it Interloper}: this is a transition from a different absorption system that happens to blend with a target line. Its species/transition may or may not be identified;
\item {\it Spectral segment}: a section of the spectrum generally flanked by absorption-free continuum regions, selected manually before running AI-VPFIT;
\end{itemize}

\subsection{Stage 1 -- Creating a preliminary model using primary transitions}\label{subsec:stage1}

In general, the dataset to be fitted will comprised multiple spectral segments, to be modelled simultaneously. By ``spectral segment'', we mean simply a short section of the spectrum that we wish to fit that is flanked by continuum regions in which no absorption is detected. In Stage 1, partially for computing speed considerations, we first build a preliminary model prior to modelling the entire dataset. The choice of the {\it primary transition(s)} is important.

A simple example might be fitting an absorption system comprised of multiple FeII transitions and the MgII 2796/2803 doublet. Let us assume, in this example, that all transitions fall longwards of the Lyman-$\alpha$ forest. If this is a damped Lyman-$\alpha$ system in which both MgII lines are mostly saturated, the sensible primary species choice is FeII.

On the other hand, for a non-DLA system, with generally lower column densities, the MgII lines may be strong but unsaturated and the FeII features may be sufficiently weak that some velocity components seen in MgII are below the detection threshold in FeII. In this case, the MgII absorption complex provides the better constraint on velocity structure and may be more suited as the primary species.

To give further examples, the requirement may be to use a single species, e.g. FeII to solve for the fine structure constant $\alpha$, or many H$_2$ lines to solve for the electron to proton mass ratio $\mu$. In these cases, the primary transition(s) are selected to be the spectral segment(s) that are likely to provide the most reliable initial model. To clarify, consider the following example. Suppose we have two FeII transitions, 1608 and 2383{\AA}, the former falling in the Lyman-$\alpha$ forest (and hence being blended with HI lines at various redshifts), the latter falling longwards of the Lyman-$\alpha$ emission line. In this situation, the FeII 2383{\AA} line should be treated as the primary transition and the FeII 1608{\AA} line as the secondary transition.

Monte Carlo methods are used to construct the preliminary model, simultaneously using all primary transitions (if more than one is being used). Initially, a single model absorption line is placed at random within the complex. Its absorption line parameters ($N$ and $b$) are default initial values, always the same, user-defined and provided by an input parameter file.

The algorithm above thus provides an initial first-guess model for the primary transitions. Non-linear least-squares, VPFIT, \citep{VPFIT,web:VPFIT}, is used to refine the initial parameters. At this point (initial iteration), we are likely to have a very poor fit to the data, using only a single profile to model an arbitrarily complex dataset. The goodness-of-fit is quantified using the corrected Akaike Information Criterion (AICc),
\begin{eqnarray} \label{eq:AICc}
AICc = \chi_p^2 + \frac{2 n k}{n-k-1} \,,
\end{eqnarray}
where $n$ and $k$ are the numbers of data points and model parameters, respectively, and
\begin{eqnarray}
\chi^2_p = \sum_{i=1}^n \frac{(F_{i,data} - F_{i,model})^2}{ \sigma^2_i}
\end{eqnarray}
where $F_{i,data}$, $F_{i,model}$, and $\sigma_i$ are the observed spectral flux, the model fit, and the estimated error on $F_{i,model}$ for the $\it i^\mathrm{th}$ pixel and the subscript $p$ indicates a summation over the data points for the primary transitions.

As is well known, the second term in Eq. \eqref{eq:AICc} penalises $\chi^2$, to prevent an arbitrarily large number of free parameters being introduced. Other statistics such as the Bayesian Information Criterion (BIC) perform a similar function but impose a stronger penalty. We defer a comparison between the various options to a forthcoming paper.

At the end of this process (Generation 1) we now have the best-fit single-profile model and can thus begin to refine the model i.e. increase its complexity until we have a statistically acceptable representation of the data. This best-fit 3-parameter model thus now becomes the parent for Generation 2, in which a second trial line is placed randomly in redshift $z$ within the fitting range\footnote{In practice, the range is defined by the lowest and highest redshift available over the multiple spectral segments used.}, assigning the trial line the same initial default parameters as before. Non-linear least-squares refinement is again carried out and AICc computed. If AICc has decreased compared to Generation 1, the current model is accepted as the Generation 2 best-fit, becoming the parent for the next generation. if AICc has {\it increased} compared to Generation 1, the model is rejected, the process repeated (i.e. a second trial line is used, again randomly placed in redshift), until AICc decreases. We call each iteration of this procedure one generation.

Iterating the above gradually increases the model complexity and improves the fit. The loop is terminated only after no reduction in AICc can be found after a user-defined number of trials (parameter $N_{line}$, set in the same initialisation file). We note that interlopers are (deliberately) not considered in this stage. Any interlopers present will be thus be automatically fitted (in this stage) as primary species. The resulting poor residuals will subsequently allow misidentifications of this sort to be corrected in Stage 3.

\subsection{Stage 2 -- Include secondary species}\label{subsec:stage2}

Stage 1 results in a good fit to the primary species alone (MgII in our example). Stage 2 comprises two steps.  Firstly, the overall dataset being fitted is increased by adding in all secondary species (FeII in our example). The absorption system structure derived in Stage 1, i.e. the best-fit set of redshifts, is replicated for each secondary species to be fitted. The initial model for the secondary species is relatively crude in that each redshift component for each of the secondary species is assigned default trial values of the column density $N$. The velocity dispersion parameters $b$ can be related either by the limiting cases of entirely {\it thermal} broadening ($b=\sqrt{2kT/m}$ such that $b_s = b_p \sqrt{m_p/m_s}$), where the subscripts $p$ and $s$ refer to primary and secondary species), entirely {\it turbulent} broadening ($b_s = b_p$), or a mixture of both, which we call {\it compound} broadening (i.e. the line broadening is defined by two parameters, the turbulent contribution and the gas temperature $T$). This is discussed in detail in Section \ref{sec:broadening}). Then, non-linear least-squares refines the parameters for all secondary species, after which AICc is again computed and stored.

If solving for either $\alpha$ or $\mu$, and if this free parameter was not introduced in Stage 1 (see earlier discussion), it can be included at this Stage. Including $\alpha$ (or $\mu$) {\it prior} to developing a complete model is important to avoid bias.

The second step within Stage 2 is to increase the model complexity, i.e. further redshift components are added, one at a time, until no descent in AICc can be found for $N_{line}$ trials. Once this happens, the Stage 2 model is complete.

\subsection{Stage 3 -- Adding interlopers}\label{subsec:stage3}

So far, we have assumed that no unidentified absorption lines are present anywhere in the data. This is generally not the case. Each quasar sightline typically intersects multiple distinct absorption systems, such that blending between lines arising in different redshift systems is common. 

There are two types of interlopers: (i) an absorption system at some other known redshift. In this case the interloper species and rest-wavelength may be identified and there may be other transitions in the same ion or from other ions at the same redshift that can be modelled simultaneously in order to best constrain the interloper parameters, or (ii) the interloper species/origin is unknown. If the interloper is identified beforehand as being of type (i), then the overall model to be fitted can include the appropriate free parameters from the outset, that is, the set-up prior to Stage 1 indicates that 2 redshift systems are to be simultaneously fitted.

If we have an interloper of type (ii), all previous (non-interloper) line parameters from Stages 1 and 2 are initially (here, in Stage 3) {\it fixed}. The reason for doing this is as follows. $\chi^2$ minimisation will act so as to `prefer' the interloper (since the interloper has no tied parameters, unlike the target line) such that the interloper may tend to replace the target line. Physically, this is obviously incorrect.

To identify possible interlopers, there are 2 potential procedures: (a) Each spectral segment is treated separately. An interloper is added in a random position within each spectral segment, one at a time, and VPFIT allowed to iterate to derive the best-fit interloper parameters (or to reject the interloper entirely), or (b) All spectral segments are dealt with at the same time i.e. one interloper is placed in each spectral segment simultaneously, with wavelength
\begin{eqnarray}
\lambda_{interloper}^i = (1-R) \times \lambda_{min}^i + R \times \lambda_{max}^i
\end{eqnarray}
where $R$ is the random seed from $0 - 1$ and $\lambda_{min (max)}^i$ is the minimum (maximum) wavelength of the $i$th segment. One could generate a different random seed for each wavelength segment but this offers no advantage. 

Superficially, (a) is the favoured option because (b) has a potential systematic trend towards over-fitting the data (i.e. fitting the data with too many absorption components). The reason for this is because introducing one single interloper at a time, and using AICc to check whether that interloper is justified, results in a final model that only contains parameters that are statistically required. On the other hand, introducing one interloper into each segment simultaneously does not permit individual AICc testing on each interloper (because of tied parameters, it is not possible to define AICc on a region-by-region basis). This argument suggests that procedure (a) is preferable in order to avoid over-fitting. However, procedure (a) is computationally very time consuming -- e.g. if we have 10 spectral segments to be fitted, the time required is 10 times as long. In terms of computing time, it is more efficient to adopt procedure (b) and subsequently test for over-fitting and repair the model where necessary (Stage 5).

\subsection{Stage 4 -- Refine and add new parameters (continuum and zero levels)}\label{subsec:stage4}

As with VPFIT, spectra are provided to AI-VPFIT either already normalised to unit continuum or with an accompanying continuum model. Since there is no guarantee the provided continuum level is always locally acceptable or optimal, VPFIT (and hence AI-VPFIT) can make linear adjustments to the continuum independently for each spectral segment. Similarly, if background subtraction during data reduction was imperfect, saturated absorption lines might in some cases be inconsistent with zero flux. The zero level can therefore be adjusted independently for each spectral segment if appropriate.

This Stage comprises two steps.\\
\noindent{\it Step 1:} Firstly, we release all the previously-fixed primary and secondary species' parameters such that they can be freely varied. Then, additional velocity components are added, one per generation. No new interlopers are included. As before, this process terminates only when no AICc descent occurs within $N_{line}$ trials.

\noindent{\it Step 2:} At this point we have only considered parameters associated with absorption lines themselves. We have not yet considered additional important parameters, such as the continuum level nor a possible zero level correction. The second step in Stage 4 is therefore to add in these additional parameters (if they are required), and refine the model again using AICc and $N_{line}$.

\subsection{Stage 5 -- Mutation; repeat earlier stages}
\label{subsec:stage5}

This Stage allows the model to evolve further by permitting the introduction of new continuum and zero-level parameters (if required). By the end of Stage 4, a good model has already been achieved. However, the only interlopers present are those introduced in Stage 3, prior to including continuum and zero-level parameters. The additional free continuum and zero-level parameters introduced in Step 2 of Stage 4 might change the model sufficiently such that further additional (or possibly fewer) interlopers are needed. Thus several procedures need repeating. Therefore we carry out the following procedures:
\vspace{-2mm}
\begin{enumerate}[leftmargin=0.5cm]
\item All continuum and zero-level parameters are temporarily fixed at the values determined at the end of Stage 4/Step 2;
\item All interlopers previously found are removed;
\item The algorithm is returned to Stage 2, but this time the Stage 2 starting model parameters (i.e. first guesses) are those from the end of Stage 4/Step 2;
\item Stages 2 to 4 are then repeated;
\item When the algorithm again reaches the end of Stage4/Step 1, the previously-fixed continuum and zero-level parameters are returned to free parameters;
\item The loop is closed by a (default or user-set) AI-VPFIT control parameter, which defines the number of repeats.
\end{enumerate}
Note that the above repeated process does not involve a return to Stage 1 because the model for the primary species has already been optimised.

We next explain the reasons behind the processes just enumerated. If, prior to Stage 5, the original continuum placement was slightly wrong, then because we may be fitting multiple transitions from a single species (e.g. several FeII transitions), a continuum error may result in a poor fit, particularly once secondary transitions are included into the modelling procedure. This point can be illustrated using simple examples. Suppose our primary transition is FeII2383 and a secondary transition is the weaker (lower oscillator strength) FeII2344. Suppose further that the original continuum placement for FeII2344 was placed slightly too low. In this example, it could easily be the case that an interloper (coincident with the FeII2344 position) provides a greater reduction in AICc rather than the (real) FeII2344 line. Such an effect clearly produces problems as follows. The lowest AICc may then be found by reducing the FeII column density so as to achieve a good fit at both lines, but an interloper is placed in the FeII2383 line to compensate for the continuum placement error. The final model then turns out wrong. Slight errors in the original continuum placement can also lead to the following problem. If, again, the original continuum had been slightly too low in places, interlopers which in reality are present in the spectrum may have been missed. Repeating Stages 2 to 4 guards against this possibility.

\subsection{Stage 6 -- Check for over-fitting and refine model parameters further}\label{subsec:stage6}

This Stage is comprised of two steps: firstly, we check the model fit to see whether all species have a consistent velocity structure and secondly (after remedying any inconsistencies) we further refine the model.

\subsubsection{Consistency check}

\noindent{\bf 1.} We first copy the best-fit velocity structure from the primary species (in Stage 5) into all secondary species. The reason for doing this is that some of the secondary species components could have fallen below a detection threshold during Stages 2, 4 and 5 and hence those parameters dropped. Clearly this makes the assumption that the primary species' velocity structure is the more reliable. Therefore, to remedy this, or at least explore whether it is avoidable, we re-insert all best-fit primary components into the secondary species. Since we re-initialise the secondary species' velocity structure (i.e. redshift parameters), we are also obliged to re-assign column density and velocity dispersion parameter values for each component, either to a set of default values or otherwise. This action necessarily pushes the current model further away from a best-fit solution. 

\noindent{\bf 2.} The second action taken is to temporarily fix any continuum and zero level parameters and redshifts (for both target lines and interlopers) to those best-fit values obtained at the end of Stage 5. The reason for doing this is as follows. By re-setting the secondary species velocity structure (and hence column densities), we temporarily create a new model that is far from a best-fit solution. Doing this can introduce a modelling instability that may occasionally generate inappropriately large parameter updates\footnote{The technical reason for this is that some parameters may move to a flat part of $\chi^2$-parameter space such that the Hessian components become very small and the search direction becomes unreliable.}.

\noindent{\bf 3.} After minimising $\chi^2$ to refine all free parameters above, the third action taken is to finally remove the temporary parameter holds described above and again to minimise $\chi^2$ to adjust all parameters.

\subsubsection{Refining}

\noindent{\bf 4.} Interlopers are introduced to the model in Stage 3. If the spectral dataset comprises multiple segments (i.e. several transitions from several different atomic species), one trial interloper is added to every segment simultaneously, the parameters are refined, and interlopers are kept if the {\it overall} AICc decreases; in fact, an interloper might increase the {\it local} AICc (i.e. for the spectral segment it lies in). For computational efficiency, to ensure that spurious interlopers do not remain in the model, this is dealt with in Stage 5 by removing each interloper one at a time and only retaining each if the global AICc does not increase.

\noindent{\bf 5.} It is possible that at this Stage the model contains regions where too many parameters have been introduced, i.e. over-fitting has occurred. There are several ways in which over-fitting could occur. One way is as follows: in Stage 1, the preliminary model is developed. If the primary species comprises only one atomic transition, there could be unidentified interlopers blended with primary species components which are therefore modelled as primary species i.e. too many parameters are attributed to the primary species. If that velocity structure is mapped into secondary species, the model as a whole becomes over-fitted. 

The second way in which over-fitting can occur is when the absorption lines associated with secondary species are weak compared to the primary species. In this situation, some secondary species' components may fall below a detection threshold, tending to increase the AICc, suggesting over-fitting in a statistical sense, even if we have physical grounds to prefer the additional parameters.

To test for and remedy both situations described above, we remove one metal component at a time, re-fit, and remove a metal component if AICc indicates a preferred model without it.

\section{Testing AI-VPFIT with synthetic spectra} \label{sec:synthetic}

A fully automated modelling process such as the one described in this paper opens new opportunities to scrutinise the limitations present in previous analyses. In particular, we do not know {\it a priori} whether the broadening of any particular absorption component is purely thermal, purely turbulent, or compound broadening. Many earlier analyses of quasar absorption systems assumed the broadening mechanism from the outset (usually turbulent) and constructed models based on that assumption. The justification for doing so was that the assumption should not bias the measured $\daa$ one way or the other. However, an important point has been overlooked; fitting the wrong model inevitably not only leads to incorrect velocity structure, but importantly, produces systematically over-fitted models. This has important and very undesirable consequences.

\subsection{Compound line broadening} \label{sec:broadening}

Consider a simple absorption system comprising only MgII2796 and FeII2383 profiles. Suppose the intrinsic gas broadening is thermal but that we use turbulent broadening to model the observed profiles. In this case, the model MgII profile will be slightly too narrow and the model FeII profile will be slightly too broad. Two mechanisms can easily compensate for this: {\it (a)} a model with 2 velocity components can match the observed profiles by adjusting the relative column densities accordingly, or {\it (b)} an interloper can be blended with the MgII profile and  $b_\mathrm{turb}$ adjusted accordingly to match the data. When $\alpha$ is an additional fitting parameter, the effects just described can add a significant systematic uncertainty. This discussion illustrates the importance of using the correct compound broadening method. Therefore, provided we have different atomic species with sufficiently different atomic masses, $b$-parameters should always be compound, i.e. they should be described by $b^2 = b^2_\mathrm{turb} + b_\mathrm{th}^2$, where $b_\mathrm{th}^2 = 2 k T/ m$, $m$ is atomic mass, and $T$ is the gas temperature. Clearly this requires a new free parameter, $T$, for each velocity component in the absorption complex.

\subsection{Generating synthetic spectra}

We base synthetic spectra on the $z_{abs}=1.15$ absorption complex in the spectrum of the bright quasar HE 0515-0414. The choice is unimportant and some other system could equally have been used. This particular absorption complex spans an unusually large redshift range, $1.14688 < z_{abs} < 1.15176$, corresponding to a velocity range $\sim 700$ km/s. It is not necessary to use such an extensive range for the purposes required here. We therefore use a subset of the system spanning the redshift range $1.14688 < z_{abs} < 1.14742$, corresponding to a velocity range $\sim 100$ km/s. For simplicity we simulate only three transitions, Mg II 2796 and 2803 {\AA} and Fe II 2383 {\AA}.

The simulated spectra are Voigt profiles convolved with a Gaussian instrumental profile with $\sigma_{res} = 1.11$ km/s. The pixel size is approximately $0.83$ km/s and the signal to noise ratio per pixel is 100 (the noise is taken to be Gaussian). The resolution and pixel size correspond to those of the HARPS instrument on the ESO 3.6m telescope \citep{Mayor2003}. The parameters used to generate the synthetic spectra are derived as follows. Firstly, we use AI-VPFIT to model the real HARPS spectrum of HE 0515-0414. The Voigt profiles used to model the real data are compound broadened. In the redshift range $1.14688 < z_{abs} < 1.14742$, we find a total of eight absorption components. Using the parameters for this model\footnote{There is one slight change: we inserted an interloper into the MgII 2796 feature to make the modelling slightly mode challenging. This can be seen in Figure \ref{fig:model_evol_stg1}.}, we create a synthetic spectrum. We refer to this synthetic spectrum as a {\it ``compound''} spectrum. The detailed parameters used to create the synthetic spectrum are given in Table~\ref{tab:syn_data}. In Section \ref{sec:Temp-importance} we describe two further synthetic spectra; one with thermal broadening and one with turbulent broadening.

\begin{table*}
\begin{tabular}{ c c c c c c c c c c c } 
\hline
 & & & & 
 \multicolumn{4}{c}{Compound} & 
 \multicolumn{1}{c}{Turbulent} & 
 \multicolumn{2}{c}{Thermal} \\ \cmidrule(lr){5-8} \cmidrule(lr){9-9} \cmidrule(lr){10-11}
\textbf{No.} &
\textbf{$\log N (\text{MgII})$} & 
\textbf{$\log N (\text{FeII})$} & 
\textbf{$\text{Redshift}$} & 
\textbf{$b_{\text{turb}}$} & 
\textbf{$T$/K} &
\textbf{$b_{\text{th}} (\textbf{Mg})$} &
\textbf{$b_{\text{th}} (\textbf{Fe})$} &
\textbf{$b(\textbf{Mg}) = b(\textbf{Fe})$} &
\textbf{$b(\textbf{Mg})$} &
\textbf{$b(\textbf{Fe})$} \\[0.5ex] 
 \hline
1 & 12.03 & 11.34 & 1.1468830 & 8.37 & 6.52E+04 & 6.68 & 4.41 & 10.71 & 10.71 & 7.07 \\
2 & 12.76 & 12.33 & 1.1469678 & 1.67 & 1.82E+04 & 3.53 & 2.33 & 3.90 & 3.90 & 2.57 \\
3 & 12.31 & 11.91 & 1.1470124 & 5.43 & 0.00E+00 & 0.00 & 0.00 & 5.43 & 5.43 & 3.58 \\
4 & 12.59 & 11.92 & 1.1471152 & 4.35 & 3.81E+04 & 5.11 & 3.37 & 6.71 & 6.71 & 4.43 \\
5 & 12.36 & 11.77 & 1.1471692 & 2.86 & 2.39E+04 & 4.04 & 2.67 & 4.95 & 4.95 & 3.27 \\
6 & 12.35 & 11.74 & 1.1472435 & 0.80 & 1.85E+04 & 3.56 & 2.35 & 3.65 & 3.65 & 2.41 \\
7 & 12.17 & 11.68 & 1.1472894 & 5.73 & 1.19E+04 & 2.85 & 1.88 & 6.40 & 6.40 & 4.23 \\
8 & 12.15 & 11.97 & 1.1474175 & 3.58 & 1.11E+04 & 2.76 & 1.82 & 4.52 & 4.52 & 2.98 \\
\hline
\textbf{Interloper} & \textbf{$\log N$} & & \textbf{$\text{Redshift}$} & \textbf{b} & \\
\cmidrule(lr){1-5}
1 & 12.00 & & 3.9381818 & 0.80 & 
\\ \cmidrule(lr){1-5}
\end{tabular}
\caption{Model parameters used to generate the three synthetic spectra discussed in Sections \ref{sec:synthetic} and \ref{sec:Temp-importance}. The model comprises eight components and one interloper. Columns 5-8 correspond to the compound synthetic spectrum, column 9 to the turbulent one, and columns 10, 11 to the thermal one. Columns 7 and 8 are the thermal $b$-parameters for MgII and FeII, i.e $\sqrt{ 2 k T/ m(\text{Mg, Fe})}$. MgII and FeII have the same $b$-parameter for turbulent broadening, as shown in column 9. Columns 10 and 11 give the thermal $b$-parameters for MgII and FeII with $b (\text{FeII}) = b (\text{MgII}) \sqrt{ m(\text{Mg})/ m(\text{Fe})}$. All $b$-parameters have units of km/s.
\label{tab:syn_data}}
\end{table*}

\subsection{AI-VPFIT models at each Stage}\label{sec:examples}

\begin{figure*}
\centering
{\includegraphics[width=0.82\linewidth]{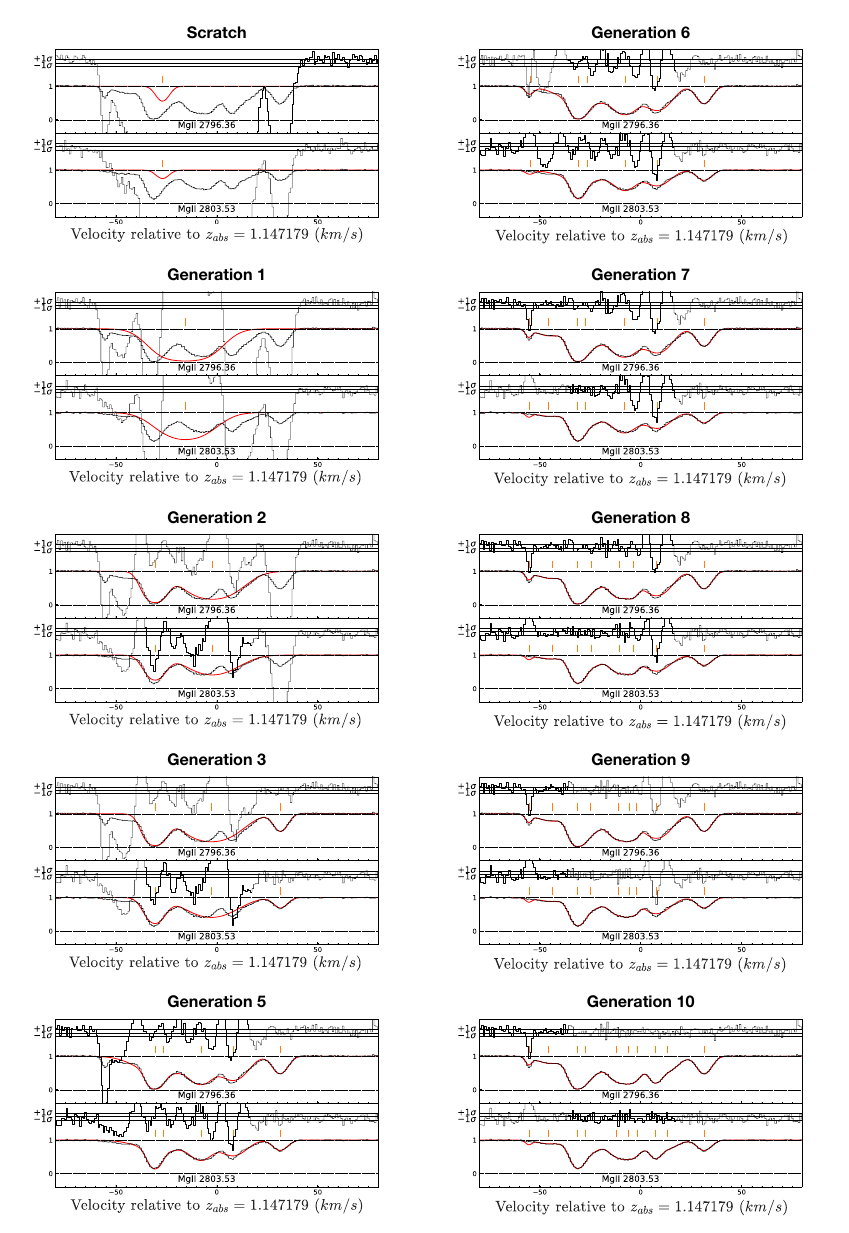}}
{\caption{Model evolution through Stage 1 of the AI-VPFIT process. The black histogram illustrates the synthetic compound spectrum. The signal-to-noise per pixel is 100. The continuous red line in the panel labelled ``Scratch'' shows the initial condition i.e. the randomly placed first-guess line. No VPFIT generation has taken place. The orange ticks shows the line centres of each metal line velocity component. The continuous red line in subsequent panels (i.e. ``Generation $\#$'') shows the best-fit compound model after each generation of VPFIT. The $\pm 1\sigma$ normalised residuals are plotted at the top of each panel. Generation 4 is missing because, by chance, the line added in that generation was rejected, such that Generations 3 and 4 are identical.}
\label{fig:model_evol_stg1}}
\end{figure*}

Fig.\ref{fig:model_evol_stg1} illustrates the evolution of the model at various points during Stage 1. Each panel shows a different generation. The initial (Scratch) model is a single randomly-placed line. At the end of Stage 1, the line has broadened and the column density increased to minimise $\chi^2$. The subsequent generations show how the model gradually evolves as more components are added. Only the primary species is shown (MgII in this case), as secondary species are not yet included. By generation 10, the model is already good, although no interlopers have been considered so the feature at $\sim -60$ km/s is (incorrectly, but knowingly) modelled as MgII. This problem is caught at Stage 3. In the example shown, AI-VPFIT continued iterating up to generation 29 but (on the basis of AICc), decided no additional velocity components were required, i.e. the model did not change for the last 19 generations.

\begin{figure*}
\centering
{\includegraphics[width=0.85\linewidth]{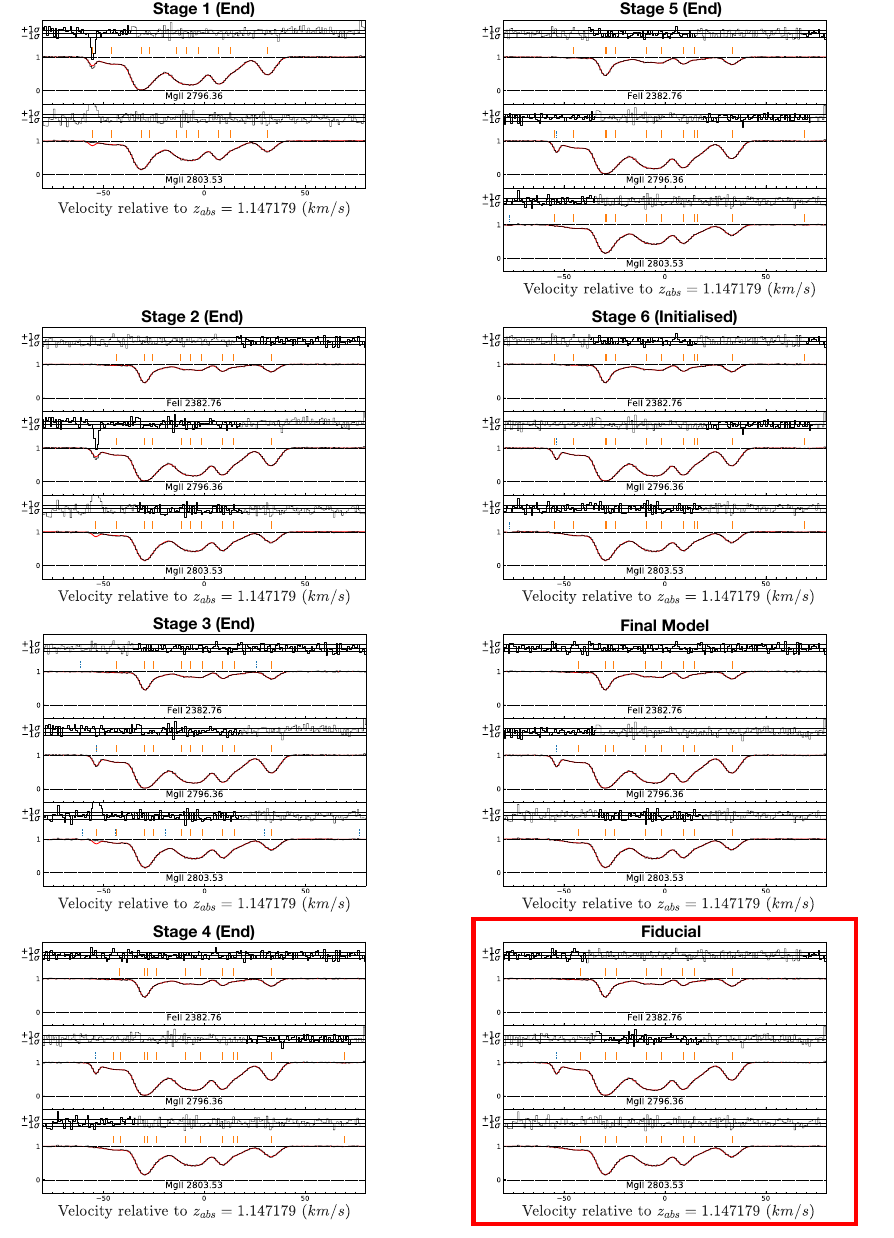}}
{\caption{Model evolution at each Stage in AI-VPFIT. 
The black histogram shows the synthetic spectrum (compound line broadening). The signal-to-noise per pixel is 100. The red continuous line illustrates the best-fit compound model. The orange/blue ticks shows the line centres of each metal line velocity component/interloper. For Stages 1 through 5, the end-model is shown. For Stage 6, we illustrate the model at the commencement of the stage, to show how the current velocity structure of the primary species (MgII in this case) is re-introduced to the secondary species (FeII), for the reasons explained in Section \ref{subsec:stage6}.}
\label{fig:model_evol}}
\end{figure*}

Fig.\ref{fig:model_evol} shows the development of the model through each Stage. For Stages 1 through 5, the model illustrated is from the end of those stages. In the example used, generation 10 is the end-point of Stage 1 so is replicated as the top-left panel in Fig.\ref{fig:model_evol}. The model at the end of Stage 1 comprises 10 components. In Stage 2, the secondary species (in this case only FeII) is included. The panel illustrating Stage 6 shows the initial condition of that Stage, where all primary species' velocity components present by the end of Stage 5 are included but {\it all} corresponding secondary species components are {\it replaced}. In the example used, 3 weak FeII components (at about $-60$, $+35$, and $+90$ km/s) are replaced as the initial condition for Stage 6. The motivation for doing so is to check that a genuine velocity component in the secondary species has not been mistakenly replaced by an interloper in Stage 5 (this is possible since the decision making is based purely on AICc).

The final best-fit velocity structure is virtually indistinguishable from input model, labelled (``Fiducial'') and highlighted with a red box to indicate this is the input and not a fitted model: the number of derived absorption components is 8 with one interloper in the MgII2796 line.

\section{The importance of the free parameter T} \label{sec:Temp-importance}

\begin{center}
\begin{figure}
\centering
{\includegraphics[width=1.0\linewidth]{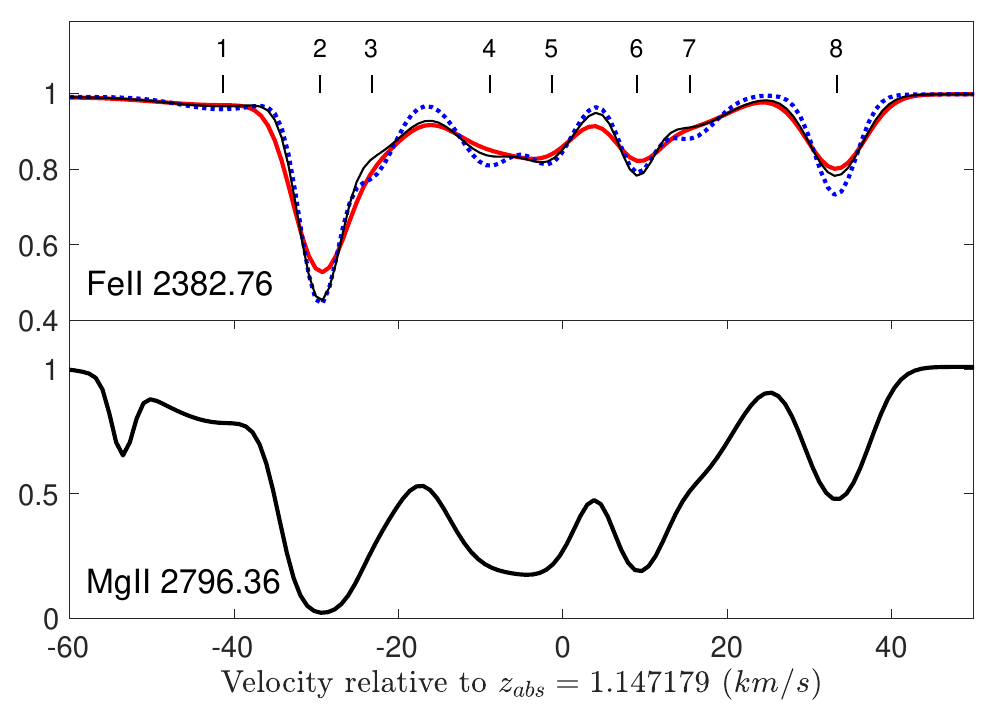}}
{\caption{Upper panel: FeII 2383 for turbulent (red continuous curve), thermal (blue dotted curve) and compound (thin black curve) broadening. Lower panel: MgII 2786, where the line broadening is $b_\mathrm{th}$(Mg) from Table \ref{tab:syn_data}. The figure illustrates the variations between each broadening model and hence highlights the importance of using the correct fitting model. For example, line 2 has a low turbulent $b$-parameter and a relatively high temperature, so the thermal and compound line centres are similar and much deeper than the turbulent line. Line 8 has a larger turbulent $b$-parameter and a lower temperature, so the turbulent and compound models are closer, the deeper line centre being the thermal model.}
\label{fig:comparison}}
\end{figure}
\end{center}

The decision as to which type of absorption line broadening to use when fitting real data is important. As alluded to earlier, previous analyses have generally assumed turbulent broadening on the basis that the assumption cannot bias estimating $\daa$. This assumption seems reasonable intuitively and is likely to be correct in a statistical sense. However, in fact it turns out that the choice of line broadening in the modelling procedure is very important.

In Section \ref{sec:synthetic}, AI-VPFIT was applied to a synthetic spectrum generated with compound broadening. This is of course not the only option. Most modelling of high resolution quasar spectra has previously been carried out assuming either turbulent or thermal broadening or both. Here we show the pitfalls of these assumptions and argue that, for reliable results, it is important to include an additional free parameter, T, for each velocity component in the model.

We begin by generating 2 additional synthetic spectra; to create the turbulent synthetic model, all parameters are kept the same except for the $b$-parameter of each absorption component, which is now set to be equal to the Mg II turbulent component. To create the thermal synthetic model, we again use the observed MgII $b$-parameters, although now the FeII $b$-parameters are adjusted according to their atomic mass. The result of the above is the creation of 3 synthetic spectra, one where the line broadening takes into account both turbulent and thermal components of the $b$-parameter, one where the broadening is assumed to be purely turbulent, and one where the broadening is assumed to be purely thermal. The detailed parameters used to create the turbulent and thermal synthetic spectra are given in Table~\ref{tab:syn_data}. Those model spectra (and the corresponding compound model), are as shown in Fig.\ref{fig:comparison}.

\begin{figure*}
\centering
{\includegraphics[width=0.99\linewidth]{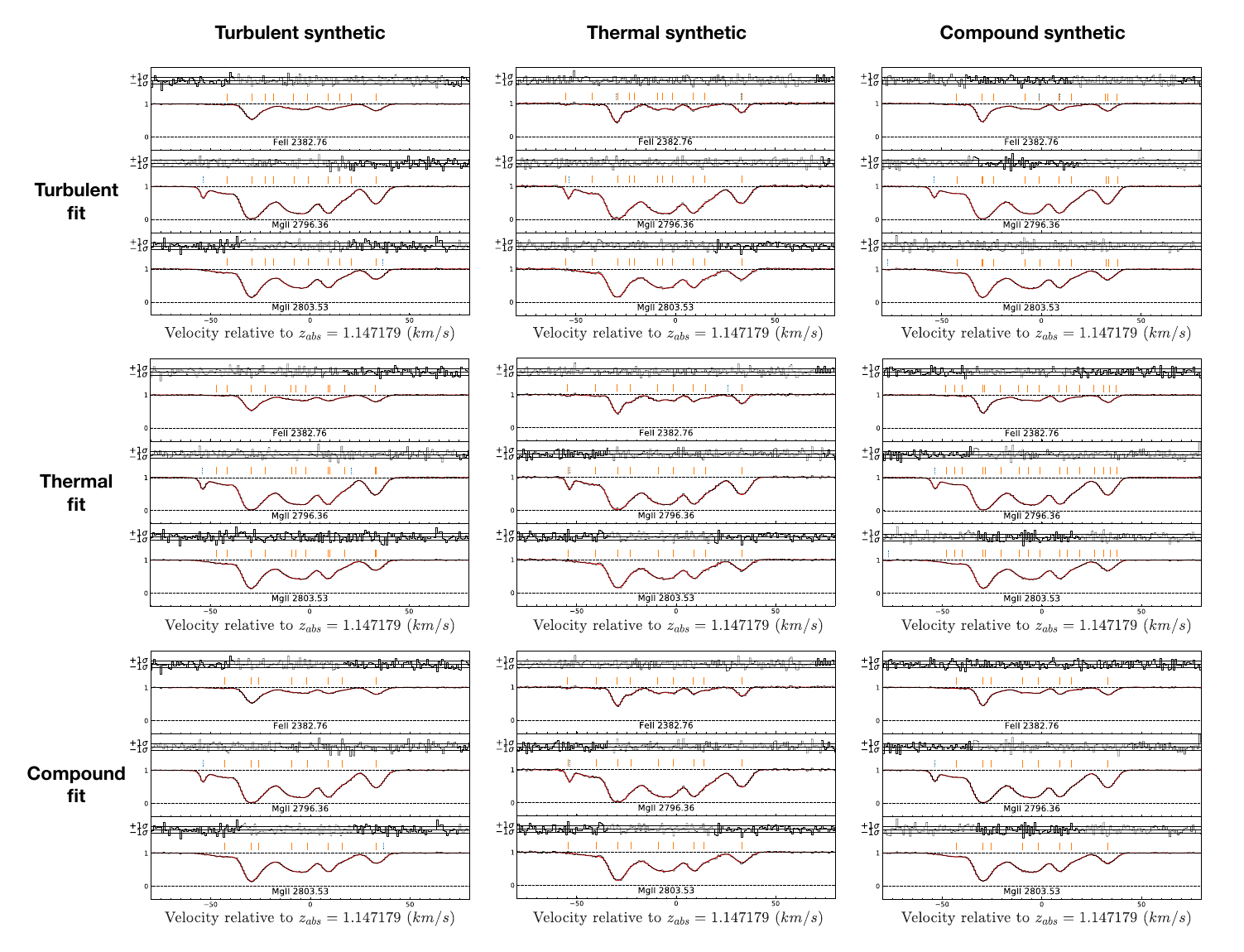}}
{\caption{Synthetic spectra and lowest AICc AI-VPFIT models. Tick marks indicate component positions. The signal to noise per pixel is 100. Other parameters used to generate the synthetic spectra are given in Section \ref{sec:synthetic}. The left hand column is for the turbulent synthetic spectrum, modelled using turbulent broadening (top), thermal broadening (middle), and compound broadening (bottom). The middle column illustrates the thermal synthetic spectrum. The right hand column illustrates the compound synthetic spectrum. Normalised residuals are plotted above each spectrum. The horizontal parallel lines indicate the 1$\sigma$ ranges.}
\label{fig:vel-structure}}
\end{figure*}

\begin{figure*}
\centering
\includegraphics[width=0.98\linewidth]{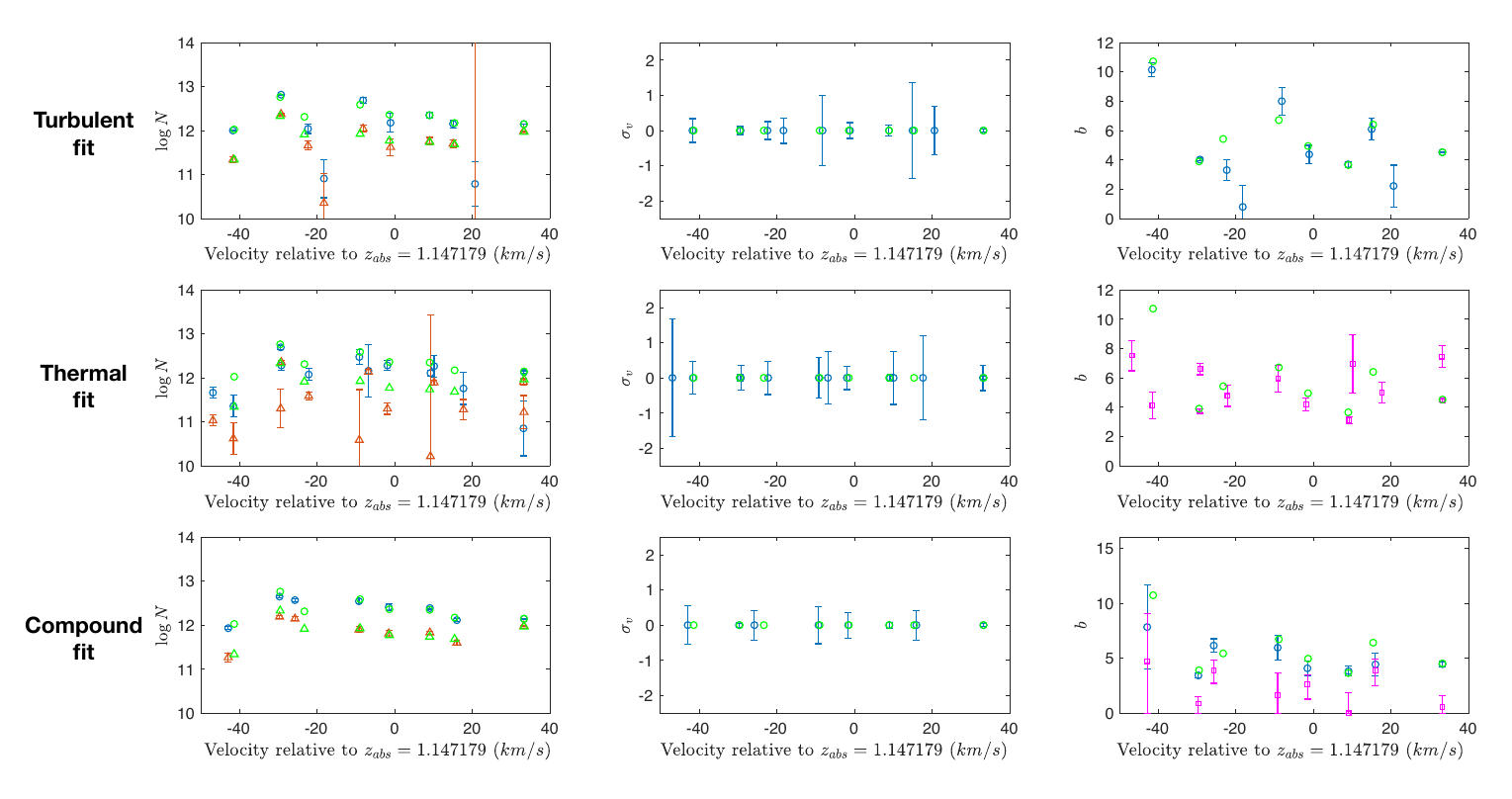}
{\caption{Results of modelling the {\it synthetic turbulent} spectrum. Each point illustrates the parameter for one velocity component. Top row: the fitted model is a turbulent model. Middle row: the fitted model is thermal. Bottom row: the fitted model is compound broadening. Green points (which has no error bars) represent the true parameter values i.e. those used to generate the synthetic spectrum. Blue points (circles) show the best-fit parameters for MgII. Red triangular points show the best-fit parameters for FeII. VPFIT error bars for each parameter are shown in all cases. Left column: $\log N$ versus velocity. Middle column: the error bars illustrate the position uncertainty for each velocity component (expressed in velocity units rather than redshift units). Each component (the hollow circle) is plotted at zero offset (along the $y$-axis) for simplicity. Since all species redshifts are tied, MgII and FeII components share the same redshifts. Right column: $b$-parameter versus velocity. Green circles (no error bars) illustrate the {\it total} $b$ value used for each MgII component, blue circles show $b_\mathrm{turb}$, and magenta squares show $b_\mathrm{th}$ for MgII. We have not illustrated $b$ values for FeII to preserve clarity.}
\label{fig:turb-stat}}
\end{figure*}

\begin{figure*}
\centering
{\includegraphics[width=0.98\linewidth]{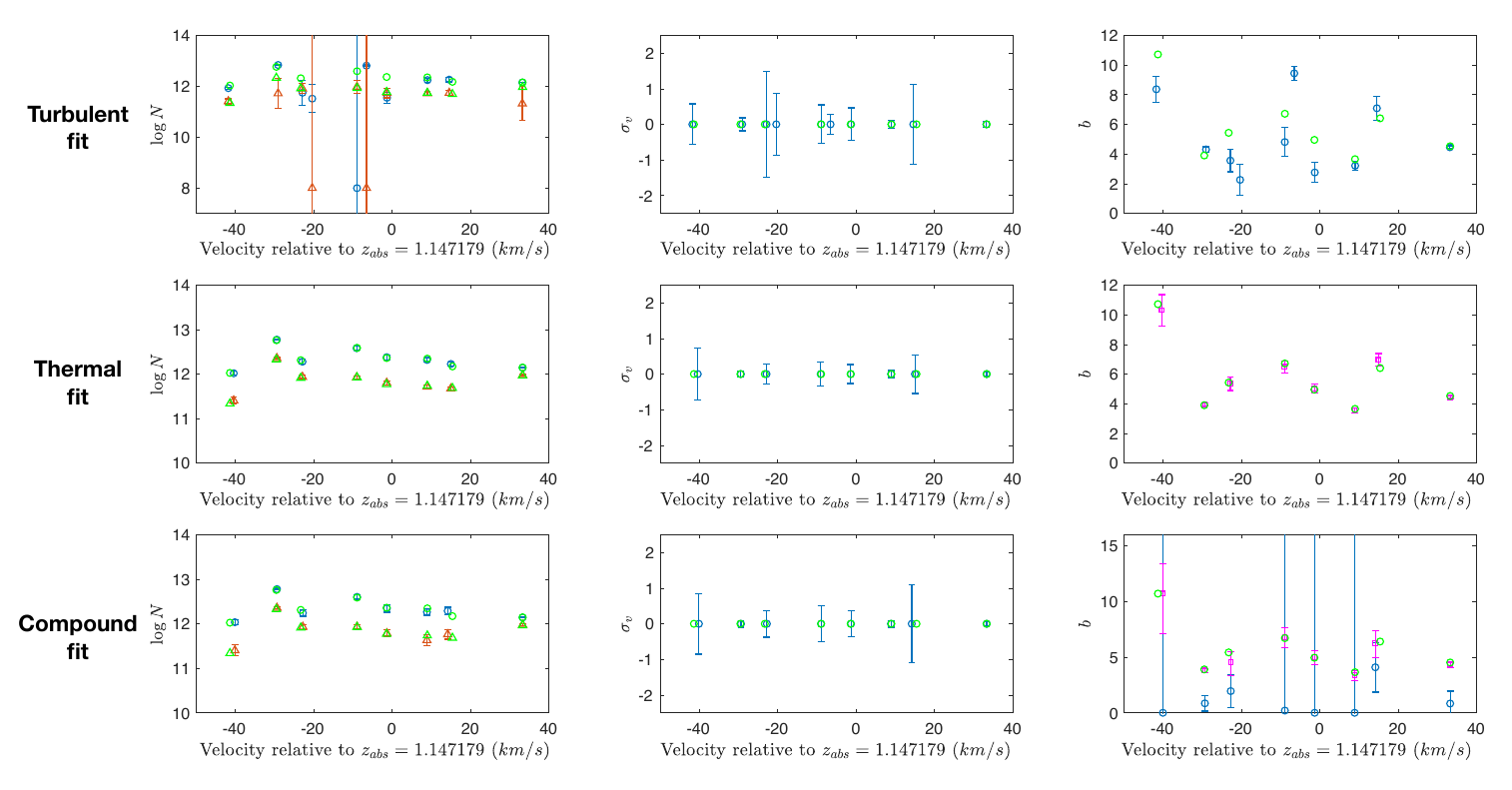}}
{\caption{Same as Fig.\ref{fig:turb-stat} except using the {\it thermal synthetic} spectrum.}
\label{fig:therm-stat}}
\end{figure*}

\begin{figure*}
\centering
{\includegraphics[width=0.98\linewidth]{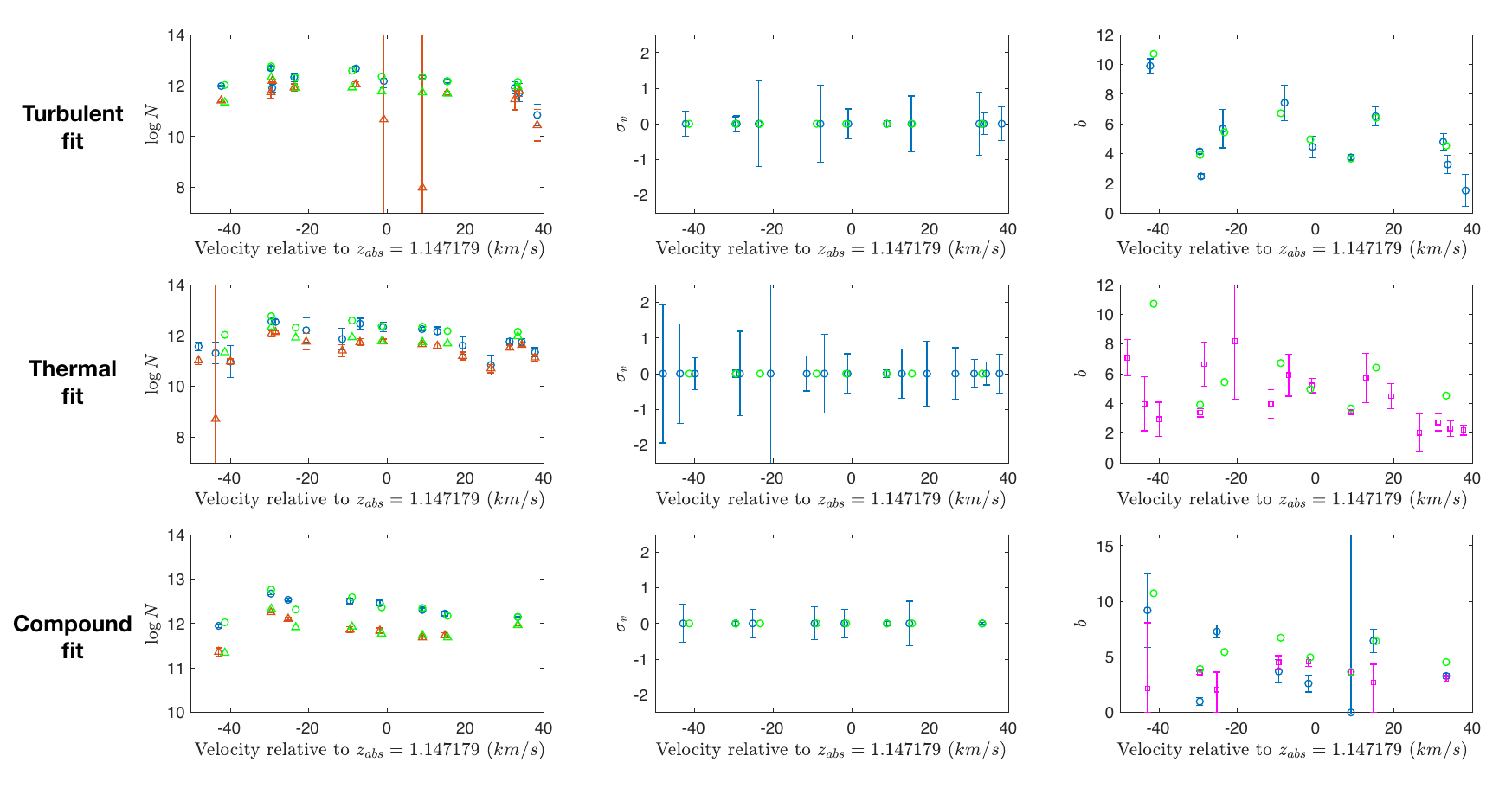}}
{\caption{Same as Fig.\ref{fig:turb-stat} except using the {\it compound synthetic} spectrum.}
\label{fig:temp-stat}}
\end{figure*}

Fig. \ref{fig:vel-structure} shows a set of nine results. Each of the three synthetic spectra (turbulent, thermal, and compound line broadening) was modelled assuming each line broadening mechanism. The spectrum and model in each panel is as described in the figure caption. The continuous orange vertical lines and blue dashed lines indicate the line centres of the target lines and interlopers. Figs. \ref{fig:turb-stat}, \ref{fig:therm-stat}, and \ref{fig:temp-stat} present the central values and errors for the $\log N$, $z$ and $b$-parameters from each best-fit model in Fig. \ref{fig:vel-structure}. These figures illustrate that modelling a turbulent spectrum with a thermal model leads to errors, and {\it vice versa}. AI-VPFIT sometimes needs two components to achieve the best-fit model for what is in reality a single component. The parameter error estimates are typically poorly determined for these ``double components''. 

Whilst ensuring that the model solves simultaneously for both thermal and turbulent components of the observed $b$-parameter, this does come with a cost, and may not always be possible. If the species fitted simultaneously (i.e. the primary and secondary species) have fairly similar atomic masses, $b_\mathrm{turb}$ and $b_\mathrm{th}$ approach degeneracy, leading to a huge uncertainty estimates on the gas temperature. In the synthetic models explored here, the only two elements ``observed'' are Mg (A=24) and Fe (A=56). Even then, slight degeneracy translates into quite large uncertainties, as is illustrated in the bottom right panel in Figs. \ref{fig:turb-stat}, \ref{fig:therm-stat}, and \ref{fig:temp-stat}.

Table \ref{tab:manyalpha} summarises the results discussed above. When the synthetic spectrum broadening and fitted model broadening match, we can see that the number of absorption components needed to satisfactorily fit the data is minimised (top black line in each section in the table). Conversely, when the ``wrong'' model is used, additional components are needed. It is now well-established that the line broadening mechanism seen in quasar absorption systems is rarely (perhaps never) entirely turbulent or entirely thermal i.e. it is important to use a compound model. Importantly, for all three synthetic spectra, a compound model performs very well in terms of finding the correct number of components (last column), as would be expected.

\begin{center}
\begin{figure}
\centering
{\includegraphics[width=1.0\linewidth]{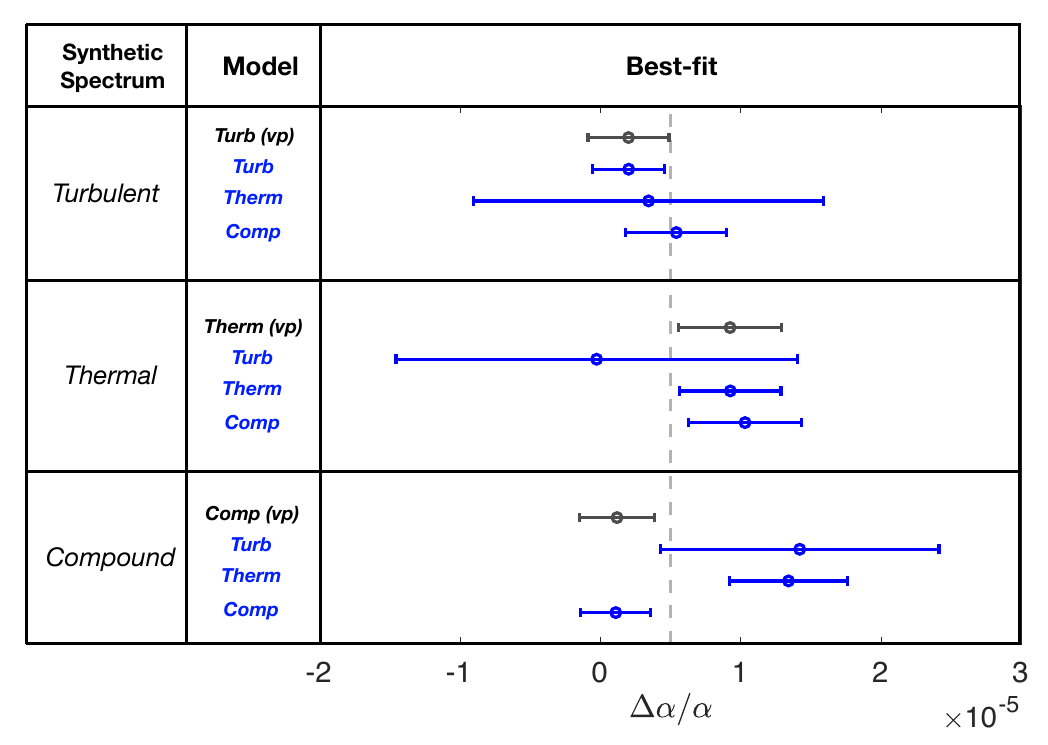}}
{\caption{Comparing the fitted $\daa$ for each permutation of synthetic spectrum and model type. The left-hand column indicates the synthetic spectrum used. The second column indicates the line broadening mechanism used in modelling. ``Turb (vp)'' means that VPFIT was used with the true model parameters as starting guesses; this point is shown in black in the third column, which also shows the best-fit $\daa$ for a turbulent, thermal, and compound model, with corresponding parameter 1$\sigma$ uncertainties (from VPFIT). The dashed line illustrates the true value.}
\label{fig:manyalpha}}
\end{figure}
\end{center}

Fig.\ref{fig:manyalpha} illustrates the results obtained from modelling the synthetic spectra described above. The figure is divided into three panels, corresponding (top to bottom) to the synthetic data being turbulent, thermal, and compound respectively. Four $\daa$ measurements are illustrated within each panel. The highest point illustrates the fitted value of $\daa$ using VPFIT only, where the starting parameter guesses were the input parameters i.e. the parameters used to generate the synthetic spectrum. The second point down illustrates the result of AI-VPFIT fitting the synthetic data using a turbulent model. The third point down illustrates a thermal model and the lowest point a compound model. The numerical results are listed in Table \ref{tab:manyalpha}. The results can be summarised as follows:
\begin{enumerate}[leftmargin=0.5cm]
\item When fitting the synthetic data with the correct model, the results are well-behaved. As expected, the fitted $\daa$ agrees well with the true value (illustrated by the vertical dashed line in Fig.\ref{fig:manyalpha}).
\item When the {\it wrong} model is used, the $\daa$ uncertainty estimate is magnified by a factor of 3 or 4. This is because imposing the wrong broadening mechanism necessarily results in additional velocity components (target lines or interlopers or both) being introduced to achieve a satisfactory fit to the data. The consequence of the additional fitting parameters is that $\daa$ is more poorly constrained. In the last case (compound synthetic data), the thermal model produces a $\daa$ estimate that is almost 2$\sigma$ away from the correct value.
\item For all three synthetic spectra, when a compound model is used, the results are good, as would be expected, since the model incorporates the full range of possibilities.
\end{enumerate}

Based on the example complex presented here, we conclude that using the wrong broadening mechanism can have highly undesirable consequences. The solution is that wherever possible (i.e. when different species with sufficiently different atomic masses are modelled simultaneously), one should use a compound model.

\begin{table}
\begin{tabular}{ l l r r c} 
\hline 
Spectrum &
Model & 
$\daa$ & 
$\sigma(\daa)$ & 
\# of lines \\[0.5ex] 
\hline
\textbf{\it Turbulent} & \textbf{\it Turb (vp)} & 1.99E-06 & 2.89E-06 & 8 + 1 \\
 & \textbf{\color{blue} \it Turb} & {\color{blue} 2.00E-06} & {\color{blue} 2.56E-06} & {\color{blue} 10 + 2} \\
 & \textbf{\color{blue} \it Therm} & {\color{blue} 3.42E-06} & {\color{blue} 1.25E-05} & {\color{blue} 13 + 2} \\
 & \textbf{\color{blue} \it Comp} & {\color{blue} 5.40E-06} & {\color{blue} 3.60E-06} & {\color{blue} 8 + 2} \\
\hline
\textbf{\it Thermal} & \textbf{\it Therm (vp)} & 9.23E-06 & 3.69E-06 & 8 + 1 \\
 & \textbf{\color{blue} \it Turb} & {\color{blue} -2.86E-07} & {\color{blue} 1.43E-05} & {\color{blue} 11 + 3} \\
 & \textbf{\color{blue} \it Therm} & {\color{blue} 9.25E-06} & {\color{blue} 3.61E-06} & {\color{blue} 9 + 2} \\
 & \textbf{\color{blue} \it Comp} & {\color{blue} 1.03E-05} & {\color{blue} 4.05E-06} & {\color{blue} 9 + 1} \\
\hline
\textbf{\it Compound} & \textbf{\it Comp (vp)} & 1.17E-06 & 2.66E-06 & 8 + 1 \\
 & \textbf{\color{blue} \it Turb} & {\color{blue} 1.42E-05} & {\color{blue} 9.94E-06} & {\color{blue} 11 + 4} \\
 & \textbf{\color{blue} \it Therm} & {\color{blue} 1.34E-05} & {\color{blue} 4.23E-06} & {\color{blue} 16 + 2} \\
 & \textbf{\color{blue} \it Comp} & {\color{blue} 1.08E-06} & {\color{blue} 2.50E-06} & {\color{blue} 8 + 1} \\
\hline
\end{tabular}
\caption{The fitted $\daa$ for each permutation of synthetic spectrum and model type - illustrated graphically in Fig.\ref{fig:manyalpha}. As an example, in the first section above, the input spectrum was generated using turbulent modelling. The first model listed as {\it Turb (vp)} means that that model also had turbulent broadening. The ``(vp)'' indicates that the initial-guess VPFIT model comprised the true (input) model parameters. See the caption to that figure for further details. The last column shows the numbers of velocity components plus the number of interlopers in the model, within the velocity range of $\pm 80$ km/s, as illustrated in Fig.\ref{fig:vel-structure}.
\label{tab:manyalpha}}
\end{table}

\section{Model non-uniqueness and impact on $\daa$}
\label{sec:nonuniqueness}

\begin{figure}
\centering
{\includegraphics[width=0.98\linewidth]{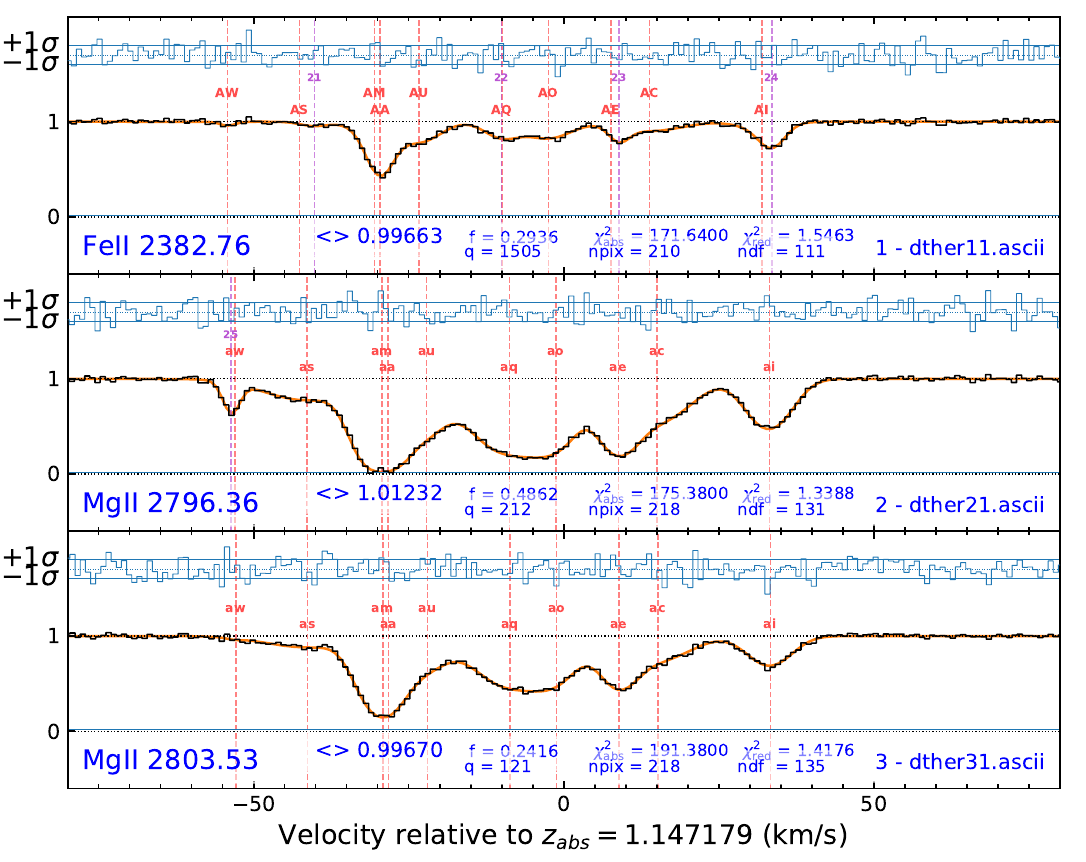}}
{\caption{Illustration of the model non-uniqueness problem. The input spectrum is thermal and the fitted model is turbulent. This model passed all acceptance tests described in Section \ref{sec:AI} and is a minimum AICc result with $\chi^2_n = 0.91$. However, it gives $\daa = 6.78 \pm 1.63 \times 10^{-5}$, compared to the ``true'' (synthetic spectrum) input value of $\daa =  5.0  \times 10^{-6}$. The discrepancy is caused by a local $\chi^2$ minimum.}
\label{fig:wrongmin}}
\end{figure}

It is important to understand that the modelling process outlined in this paper, in the context of solving for $\daa$, is {\it not} the same as the interactive process followed by a human being. In previous studies, where no AI methodology has been used, and where a human has interactively constructed an absorption line model, this is done by assuming throughout that $\daa=0$. Once a good model has been found, the procedure generally adopted in previous analyses has been to only then include $\daa$ as a free parameter. 

We show in this Section that in fact very good models can be (and are) found with disparate values of $\daa$. Given enough computations, different starting points (equivalent to using different random seeds in the AI-VPFIT code) may reach different solutions. Whilst this is a well-known characteristic of non-linear least-squares procedures applied to complex datasets, it has not been studied in any detail in the context of modelling high-resolution quasar absorption spectra.

Fig.\ref{fig:wrongmin} shows an AI-VPFIT model, obtained using the thermal synthetic spectrum fitted using a turbulent model. The model illustrated gives good normalised residuals and appears to be a good representation of the data. However, the synthetic spectrum was generated using $\daa = 5 \times 10^{-6}$, yet the best-fit value found for this particular turbulent fit is $\daa = 68 \pm 16 \times 10^{-6}$. The model is thus 4$\sigma$ away from the ``true'' value in this case. Inspecting the model details reveals how this occurs. Referring to Fig.\ref{fig:wrongmin}, the FeII column density of velocity component AI has decreased relative to its ``true'' value, to permit the interloper 24. The same effect occurs at velocity component AE. The interloper 25 is actually present in the data. The strong line under the AM and AA components is actually a single component. However, AI-VPFIT has introduced two components here, such that the column densities of the FeII AA and MgII AM component dominate the line centroid positions. The ``true'' model in fact comprises eight velocity components and one interloper (25). However, Fig.\ref{fig:wrongmin} comprises ten target lines and five interlopers.

In this example, clearly a false minimum in $\chi^2$ has been found. There is no {\it a priori} reason to reject this model. The only potentially tell-tale signs of there being a problem with this model are (i) an apparent excess of close blends in the stronger components, and (ii) the value of $\daa$ obtained is inconsistent with that obtained using the compound and thermal models shown in Fig.\ref{fig:vel-structure}.

The effect described above, i.e. finding a statistically acceptable model associated with a false minimum, is a natural consequence of an unbiased modelling the process. (An interactive i.e. human method may artificially avoid the problem by fixing $\daa = 0$ throughout the model-building process). The correct solution is to repeat the whole fitting process at least twice (and preferably more), searching for additional solutions with fewer velocity components and a lower AICc. We also note that, from calculations done so far, the number of acceptable models (i.e. the degree of non-uniqueness) is likely to be reduced when fitting a compound model (as would be expected).

\section{Overfitting - AIC and BIC} \label{sec:Overfitting}

In \cite{Bainbridge2017,gvpfit17} and in the present paper we have made extensive use of the AICc statistic \citep{Hurvich1989} to select candidate models. In the course of this work, we also compared AICc results with those using the Bayesian Information Criterion, BIC \citep{Bozdogan1987}. Both statistics add a penalty to the usual $\chi^2$ statistic, the value of which increases with increasing number of model parameters. The purpose is to end up with a final model having the ``right'' number of parameters. In other words, both AICc and BIC try to limit the model complexity to avoid ``over-fitting''. The AICc (Eq.\eqref{eq:AICc}) and BIC penalties are
\begin{equation} \label{eq:penalties}
2nk/(n-k-1) \ \ \mathrm{and} \ \ k \log n \,,
\end{equation}
where $n$, $k$ are the number of data points and free parameters respectively.

\begin{figure}
\centering
{\includegraphics[width=0.98\linewidth]{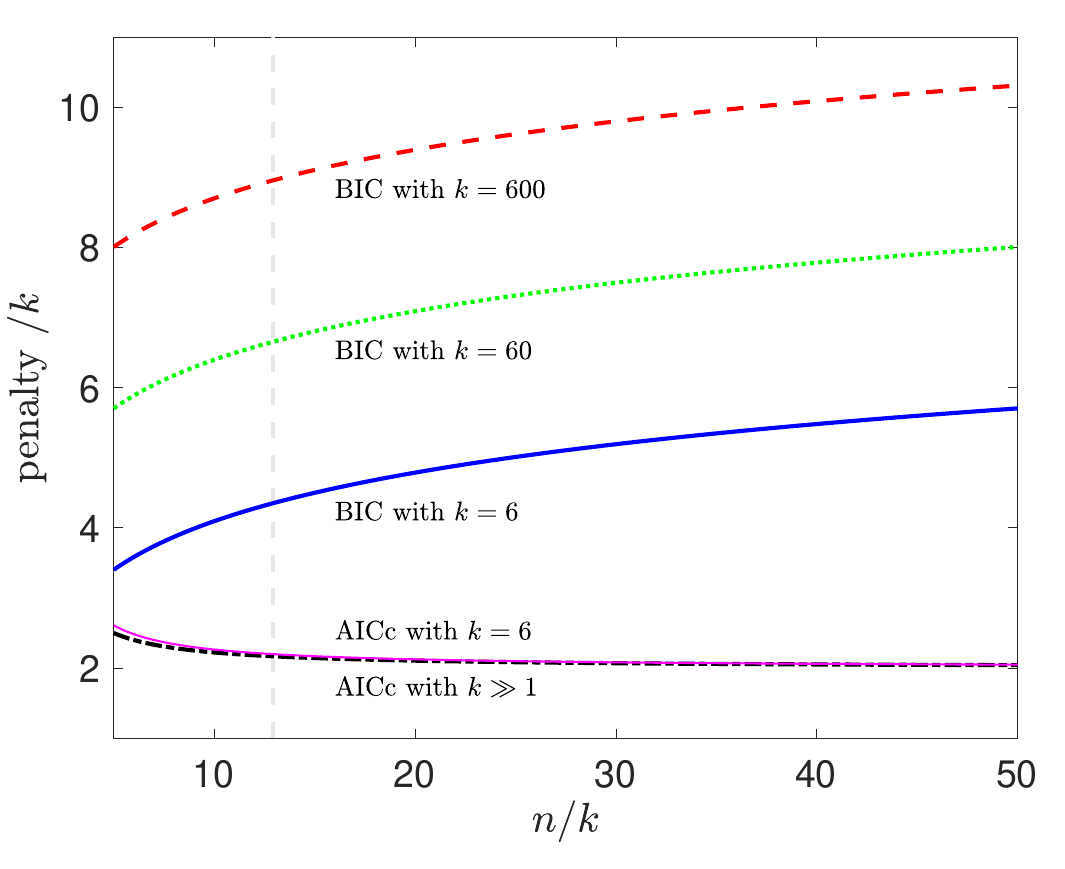}}
{\caption{Penalty per free parameter (Eq.\ref{eq:penalties} divided by $k$) vs. the number of pixels per parameter, $n/k$. The blue (solid), green (dotted), and red (dashed) curves illustrate the BIC penalty term for different $k$. At fixed $n/k$, BIC increases relatively rapidly as more model parameters are introduced. At fixed $n/k$, BIC also increases as the spectral fitting region $n$ is increased. The mageneta (continuous) and black (dot-dash) curves shows the AICc penalty term. The AICc penalty is insensitive to $k$ and relatively insensitive to $n/k$. The vertical grey (dashed) line indicates $n/k = 646/50$ for the synthetic model of Fig.\ref{fig:comparison}. }
\label{fig:penalties}}
\end{figure}

Fig.\ref{fig:penalties} illustrates the contribution to the penalty from each free parameter in the model. Placing the spectral fitting boundaries (i.e. defining $n$) is to some extent arbitrary. Thus in this application to absorption line spectroscopy, a penalty that is relatively insensitive to $n$ is preferable. However, including pixels far from line centre is required if we want to fit continuum parameters. These two considerations alone show that AICc is preferable to BIC in our application.

Nevertheless, even though we have adopted AICc in this paper, it is not entirely suitable. The spectral fitting boundaries influence the AICc penalty
and the former are ill-defined. Moreover, for an unsaturated absorption line (for example), pixels far from the line centre impact little on the line profile whilst the converse is true, yet each pixel across the absorption line or complex carries the same weight in the AICc penalty term. Another way of expressing this is that each parameter can influence specific regions within the dataset (but have little or no impact elsewhere); the parameters describing an absorption component at $-$50 km/s in Fig.\ref{fig:comparison}, for example, have virtually no influence on the model intensity at +50 km/s. The AICc (and BIC) fail to take such an effect into account, hence these ICs are not optimal for absorption line spectroscopy and some other (new) IC statistic is needed \citep{webb2020getting}.

Finally, the details of model non-uniqueness (Section \ref{sec:nonuniqueness}) are likely to depend on the statistic used i.e. AICc or something else. These considerations are beyond the scope of the present paper and will be addressed in subsequent work.

\section{Discussion and conclusions}\label{sec:conclusions}

In this paper we have extended previous work that  combined a genetic algorithm with non-linear least-squares to automatically model absorption line data. ``Interactive" absorption line fitting is subjective and generally not reproducible. The method presented here brings objectivity and repeatability, important requirements for the increasingly high-quality data from new and forthcoming spectroscopic facilities like ESPRESSO/VLT and HIRES/ELT \citep{2020arXiv201112317M}. 

An AI method provides an unbiased estimate of $\daa$. The same is not true of any interactive method that holds $\daa=0$ during the model construction phase, only to ``switch on'' $\daa$ as a free parameter at the end, once the model is essentially complete. A procedure of this sort preferentially selects local minima closest to $\daa=0$. 

Synthetic spectra, based on a well-known intermediate redshift absorption system, have been created and used to test the method's performance. We have shown that model development goes wrong i.e the wrong velocity structure is obtained, if the wrong broadening mechanism is used to model the observational data. High quality observations of quasar absorption systems show that in general the line broadening does indeed contain both thermal and turbulent contributions simultaneously. However, much of the previous work on estimating $\daa$ has been done assuming a single broadening mechanism. We have shown, through example, that there are highly undesirable consequences of this: the wrong velocity structure may easily be obtained, resulting in far less reliable parameter estimates. Fig. \ref{fig:manyalpha} shows this clearly for $\daa$.

The extent to which this model non-uniqueness issue applies even when the {\it correct} model is used is not yet clear. It is possible, even likely, that $\chi^2$--parameter space is sufficiently complex, with multiple local minima, that false minima result even when solving for $b_\mathrm{th}$ and $b_\mathrm{turb}$ simultaneously. If so, model non-uniqueness limits the precision achievable from any single measurement i.e. it is not possible to reach the theoretical statistical limit using a single observation of $\daa$, no matter how precise the wavelength calibration. The implication is that, by nature, the problem of measuring $\daa$ is a statistical one, requiring a large sample of measurements to render the non-uniqueness systematic negligible. This principle should perhaps underpin all future scientific efforts to determine whether the fine structure constant varies in time or space.

\section*{Acknowledgements}
We are grateful for supercomputer time using OzSTAR at the Centre for Astrophysics and Supercomputing at Swinburne University of Technology. CCL thanks the Royal Society for a Newton International Fellowship during the early stages of this work. JKW thanks the John Templeton Foundation, the Department of Applied Mathematics and Theoretical Physics and the Institute of Astronomy at Cambridge University for hospitality and support, and Clare Hall for a Visiting Fellowship during this work.

\section*{Data Availability}
The numerical simulations in this paper are based on observations collected at the European Southern Observatory under ESO programme 102.A-0697(A). The synthetic spectra themselves can be obtained from the authors on request.

\bibliographystyle{mnras}
\bibliography{mybibliography}

\bsp
\label{lastpage}
\end{document}